\documentclass[10pt,letterpaper]{article}
\usepackage{opex3}

\begin{document}

\title{Quantum theory of a spaser-based nanolaser}

\author{Vladimir M. Parfenyev and Sergey S. Vergeles}

\address{Moscow Institute of Physics and Technology, Dolgoprudnyj, Institutskij lane 9,\\Moscow Region 141700, Russia}
\address{Landau Institute for Theoretical Physics RAS, Kosygina 2, Moscow 119334, Russia}

\email{parfenius@gmail.com} 



\begin{abstract}
We present a quantum theory of a spaser-based nanolaser, under the bad-cavity approximation. We find first- and second-order correlation functions $g^{(1)} (\tau)$ and $g^{(2)}(\tau)$ below and above the generation threshold, and obtain the average number of plasmons in the cavity. The latter is shown to be of the order of unity near the generation threshold, where the spectral line narrows considerably. In this case the coherence is preserved in a state of active atoms in contradiction to the good-cavity lasers, where the coherence is preserved in a state of photons. The damped oscillations in $g^{(2)}(\tau)$ above the generation threshold indicate the unusual character of amplitude fluctuations of polarization and population, which become interconnected in this case. Obtained results allow to understand the fundamental principles of operation of nanolasers.
\end{abstract}



\section{Introduction}

In the last decade nanoplasmonics led to the emergence of many promising
applications~\cite{Stockman2011d}. One of them is a near-field generator of
nanolocalized coherent optical fields --- spaser-based nanolaser or SPASER (surface
plasmon amplification by stimulated emission of radiation), which was shown
to be an optical counterpart of the MOSFET (metal-oxide-semiconductor field
effect transistor)~\cite{Stockman2010}. The device was proposed by D.~Bergman
and M.~Stockman in the paper~\cite{Bergman2003a}. Operation principles of the
spaser-based nanolaser are similar to operation principles of the usual laser, but instead of
photons we deal with surface plasmons (SPs). The first experimental
observations were made by M.~Noginov's group~\cite{Noginov2009} and
X.~Zhang's group~\cite{Oulton2009a} and they are dated back to 2009 year.

Since the spaser-based nanolaser is a source of coherent light, it has a narrow spectral line width
above the generation threshold. There are two main possible mechanisms which
lead to the spectrum narrowing. For high Q-factor resonators, the narrowing
is determined by the domination of stimulated emission of radiation and thus, by the large
number of excited quanta in the resonator \cite{Scully, Carmicael_1}.
This situation is typical for usual lasers.
The opposite limit of low Q-factor resonators and relatively small excited quanta
corresponds to spaser operation \cite{Noginov2009}. In the case, the
narrowing can be determined by large number of the excited atoms in the gain
media and their coherence, which is achieved by the mutual
interaction of atoms through the resonator mode \cite{Bohnet2012a}. Here we show that the spaser-based nanolaser
can produce narrow spectrum of generation even if the mean number of excited
quanta in resonator is of the order or less than unity, and find the spectrum of the
generation and its statistics.

The exact analytical treatment of lasing involves quantum fluctuations, which
are responsible for finite width of the spectral line~\cite{Scully, Carmicael_1}. Description of the laser with Maxwell-Bloch
equations, see e.g. \cite{Stockman2010, Parfenyev2012}, corresponds to
mean-field approximation both for the resonator mode and
atoms. We develop a theory which allows to account for quantum
fluctuations in a low $Q$-factor resonator with the arbitrary number of
quanta both below and above the generation threshold, which interacts with
ensemble of $N$ identical active atoms.



To solve the problem completely analytically we impose some
restrictions. First, we assume that the cavity decay rate $\kappa$ is the
fastest rate in the system. Thus, the resonator mode can be
adiabatically eliminated~\cite{Cirac1992}, and the state of the spaser-based
nanolaser is fully characterized by the state of $N$ identical two-level
active atoms. Second, we believe $N \gg 1$ and thereby the fluctuations of
the state of atoms can be considered is the small-noise limit~\cite[ch.5.1.3]{Carmicael_1}.
Note that due to adiabatic mode elimination we can only resolve times $\tau \gg 1/\kappa$.  The smaller times were
considered in the paper \cite{Andrianov2013}, for the model with single active atom,
$N=1$, but only below the generation threshold, when mean number of quanta in the
resonator is well below unity.

As a result, we describe the behaviour of the spaser-based nanolaser below and above
generation threshold, and demonstrate that the spectral line narrows considerably,
when passing through the threshold. We find
the average number of plasmons in the cavity and show that this number near
the generation threshold can be of the order of unity. In the case the
coherence is preserved in a state of active atoms, which relax slowly than
$1/\kappa$. This fact fundamentally distinguishes the behaviour of the
bad-cavity nanolasers in comparison with the good-cavity lasers, where the coherence
is preserved in a state of photons. The evaluation for the number of plasmons
is in accordance with the experimental observations~\cite{Noginov2009}. Moreover, we
obtain second-order correlation function $g^{(2)}(\tau)$, $\tau \gg 1/\kappa$, and find that above the generation
threshold the amplitude fluctuations of polarization of active atoms lead to the damped oscillations in
$g^{(2)}(\tau)$. A similar dependence was observed in numerical simulations
in the paper~\cite{Temnov2009a}, and it is usual for bad-cavity lasers~\cite{Gnutzmann1998}.
However, in the case of good-cavity lasers there is no oscillations in second-order correlation function~\cite{Carmicael_1}.
We assume that the shape of the curve $g^{(2)}(\tau)$ indicates the mechanism
of the spectral line narrowing, and therefore we investigate at what relationship between cavity decay rate $\kappa$ and homogeneous broadening of active atoms $\Gamma$ the oscillations occur.
We think that obtained results are important for understanding the fundamental
principles of operation of spaser-based nanolasers.

\section{Physical Model and Methods}

We consider $N \gg 1$ identical two-level active atoms with resonant
frequency $\omega$ coupled to single strongly damped cavity, with a short
plasmon lifetime $(2 \kappa)^{-1}$ centered at the same frequency. The
interaction between the atoms and the field is described by the
Tavis-Cummings Hamiltonian~\cite{Carmicael_1}, $H_{AF} = i \hbar g (a^+ J_- -
a J_+)$, where $g$ is coupling constant, $a^+$ and $a$ are the creation and
annihilation operators of plasmons in the cavity mode \cite{Andrianov2013a}, and $J_{\alpha} =
\sum_{j=1}^N \sigma_{j \alpha}$ is collective atomic operators, where
$\sigma_{j\alpha}$, $\alpha=\{x,y,z\}$ -- Pauly matrices, and
$\sigma_{j\pm}=(\sigma_{jx}\pm i\sigma_{jy})/2$. In a bad-cavity
limit~\cite{Cirac1992} the cavity mode can be adiabatically eliminated and
the following master equation in the Schr\"{o}dinger picture for the atomic
density operator $\rho = \mathrm{tr}_{F} \rho_{AF}$:
\begin{eqnarray}
\nonumber
&\displaystyle \dot{\rho} = -i \frac{1}{2} \omega [J_z, \rho] + \frac{\gamma_{\uparrow}}{2} \left( \sum_{j=1}^N 2 \sigma_{j+} \rho \sigma_{j-} + \frac{1}{2} J_z \rho + \frac{1}{2} \rho J_z - N \rho \right) + &\\
\nonumber
&\displaystyle +\frac{\gamma_{\downarrow}}{2} \left( \sum_{j=1}^N 2 \sigma_{j-} \rho \sigma_{j+} - \frac{1}{2} J_z \rho - \frac{1}{2} \rho J_z - N \rho \right) + \frac{\gamma_{p}}{2} \left( \sum_{j=1}^N \sigma_{jz} \rho \sigma_{jz} -N \rho \right)+&\\
\label{eq:1}
&\displaystyle  + \frac{g^2}{\kappa} \left(2 J_- \rho J_+ - J_+ J_- \rho -  \rho J_+ J_- \right),&
\end{eqnarray}
where the trace is taken over the field variables. Here the active atoms are
incoherently pumped with rate $\gamma_{\uparrow}$ and we take into account
the spontaneous emission with rate $\gamma_{\downarrow}$. The
dephasing processes, which are caused mostly by the interaction with phonons,
has rate $\gamma_p$. The last term describes interaction of active atoms
through the cavity mode. Adiabatic mode elimination can be performed only if
$\kappa \gg N g^2 / \kappa, \gamma_{p}, \gamma_{\uparrow},
\gamma_{\downarrow}$. Note, that normal-ordered field operator averages can be restored by formal
substitutions $a^+(t) \rightarrow (g/\kappa)J_+(t)$ and $a(t) \rightarrow
(g/\kappa) J_-(t)$, \cite[(13.60)]{Carmicael_2}.
The same model, but with an arbitrary number $N$ of
active atoms, was considered numerically by V.~Temnov and U.~Woggon in papers
\cite{Temnov2009a, Temnov2005}, where they showed that the last term in equation
(\ref{eq:1}) leads to the cooperative effects.

Our final interest is the state of the resonator, thus it is enough to
describe the system of atoms in terms of collective atomic operators, despite
the fact that some terms in (\ref{eq:1}) cannot be rewritten in terms of
$J_+, J_z, J_-$. First, we define characteristic function
\begin{equation}\label{eq:2}
\displaystyle \chi_N (\xi, \xi^*, \eta) \equiv \mathrm{tr} \left(\rho e^{i \xi^* J_+} e^{i \eta J_z} e^{i \xi J_-} \right),
\end{equation}
which determines all normal-ordered operator averages in usual way. Next, we
introduce the Glauber-Sudarshan $P$-representation $\tilde{P}(v,v^*,m)$ as the
Fourier transform of $\chi_N (\xi, \xi^*, \eta)$, which can be interpreted as
distribution function and allows to calculate normal-ordered operator
averages as in statistical mechanics \cite[(6.118a)]{Carmicael_1}. Evolution
equation on $\tilde{P}$ can be obtained by differentiation (\ref{eq:2}) with
respect to time and replacement $\dot{\rho}$ with (\ref{eq:1}). The
results is
\begin{equation}\label{eq:3}
\displaystyle \frac{\partial \tilde{P}}{\partial t} = L \left(v,v^*,m, \frac{\partial}{\partial v},\frac{\partial}{\partial v^*},\frac{\partial}{\partial m} \right) \tilde{P},
\end{equation}
where
\begin{eqnarray}
\nonumber
&\displaystyle L = \frac{\gamma_{\uparrow}}{2} \left[ \left( e^{-2 \frac{\partial}{\partial m}} -1 \right) (N-m) + \frac{\partial^4}{\partial v^2 \partial v^{*2}} e^{2 \frac{\partial}{\partial m}} (N+m) + 2N \frac{\partial^2}{\partial v \partial v^*} \right]+&\\
\nonumber
&\displaystyle + \frac{\gamma_{\uparrow}}{2} \left( 2e^{-2 \frac{\partial}{\partial m}} -1 + 2 \frac{\partial^2}{\partial v \partial v^*} \right) \left( \frac{\partial}{\partial v} v + \frac{\partial}{\partial v^*} v^* \right)+&\\
\nonumber
&\displaystyle+ \frac{\gamma_{\downarrow}}{2} \left[ \left( e^{2 \frac{\partial}{\partial m}}-1 \right) (N+m) +  \frac{\partial}{\partial v} v + \frac{\partial}{\partial v^*} v^* \right]+&\\
\nonumber
&\displaystyle+ \gamma_{p} \left[ \frac{\partial}{\partial v} v + \frac{\partial}{\partial v^*} v^* + \frac{\partial^2}{\partial v \partial v^*} e^{2 \frac{\partial}{\partial m}} (N+m) \right]+ i \omega \left[ \frac{\partial}{\partial v} v - \frac{\partial}{\partial v^*} v^* \right]+&\\
\nonumber
&\displaystyle+ \frac{g^2}{\kappa} \left[ 2 \left( 1 - e^{-2\frac{\partial}{\partial m}} \right) v v^* - \left(  \frac{\partial}{\partial v} vm + \frac{\partial}{\partial v^*} v^*m \right) + \frac{\partial^2}{\partial v^2} v^2 +\frac{\partial^2}{\partial v^{*2}} v^{*2}\right].&
\end{eqnarray}
The closed form of the equation confirms the possibility of describing the
system in terms of collective atomic operators.

The exact solution of the equation (\ref{eq:3}) is strictly singular due to
the exponential factors in $L$, which describe transitions in active atoms.
Moreover, the solution cannot be found in analytical form. However, we can
obtain an approximate nonsingular distribution, replacing (\ref{eq:3}) by a
Fokker-Planck equation. The key element to such replacement is a
system size expansion procedure~\cite[ch. 5.1.3]{Carmicael_1}. The
large system size parameter in our case is the number $N \gg 1$ of active
atoms. Following the system size expansion method we will obtain an adequate
treatment of quantum fluctuations in the first order in $1/N$.

To make a systematic expansion of the phase-space equation of motion in
$1/N$, we move into rotating frame and introduce dimensionless polarization
per one atom $\sigma=\mathop{\mathrm{tr}}[\rho {J}_{-}e^{i \omega t}]/N$ and
inverse population $n =
 \mathop{\mathrm{tr}}[\rho J_{z}]/N $. Next, we separate the mean values ​​and fluctuations in the
phase-space variables
\begin{equation}\label{eq:4}
\displaystyle v e^{i \omega t}/N = \sigma + N^{-1/2} \nu,\quad m/N = n + N^{-1/2} \mu.
\end{equation}
and introduce a distribution function $\displaystyle P (\nu,
\nu^*, \mu, t) \equiv N^{3/2} \tilde{P} \left(v(\nu, t), v^*( \nu^*, t), m( \mu, t), t
\right)$, which depends on variables, corresponding to fluctuations. Using
the equation (\ref{eq:3}) and neglecting terms $\sim O(N^{-1/2})$, we obtain the
equation for a scaled distribution function $P$. More accurately, we
obtain two sets of equations: the first set describes dynamics of macroscopic
variables, and one more equation characterizes fluctuations.

\section{Macroscopic Equations and Generation Threshold}

First, we analyze the system of equations describing the dynamics of macroscopic variables, which takes a form
\begin{eqnarray}\label{eq:5}
&\displaystyle \frac{d (\sigma/n_s)}{\Gamma dt} = -  \left( 1 - \wp \frac{n}{n_s} \right) \sigma/n_s,&\\
\label{eq:7}
&\displaystyle \frac{d (n/n_s)}{\Gamma dt} = -\frac{n/n_s-1}{\Gamma T_1} - 4\wp  |\sigma/n_s|^2,&
\end{eqnarray}
where we introduce population relaxation time $T_1 =
1/(\gamma_{\uparrow}+\gamma_{\downarrow})$, homogeneous broadening $\Gamma =
\gamma_{p}+1/(2 T_1)$ and equilibrium inverse population $n_s =
(\gamma_{\uparrow}- \gamma_{\downarrow})/(\gamma_{\uparrow}+
\gamma_{\downarrow})$. The system has two different stable steady-states
solutions, depending on \textit{pump-}parameter $\wp = \wp_0 n_s$, where
$\wp_0 = Ng^2/(\kappa \Gamma)$.

In the case $\wp < 1$, one obtain the solution $n = n_s, \; \sigma = 0$. Thus,
there is no macroscopic polarization and this situation corresponds to the
nanolaser operating below generation threshold. In the opposite case $\wp >
1$, the solution takes a form $n  = 1/ \wp_0, \; |\sigma|=
(n_s/2\wp)\sqrt{(\wp-1)/\Gamma T_1}$ and it corresponds to the spaser-based nanolaser operating
above generation threshold. Overall, the picture is completely analogous to
the good-cavity laser \cite[ch. 8.1.2]{Carmicael_1}.

Note, that in the case $\wp=1$ both solutions are the same. This point
corresponds to the spaser generation threshold, which was obtained in earlier
semiclassical papers, e.g. \cite{Stockman2010, Parfenyev2012}.

\section{Quantum Fluctuations Below Threshold}

Second, we analyze the equation, which provides a linearized treatment of
fluctuations about solution to the system of macroscopic equations. In the
case below generation threshold, i.e. $\wp < 1$, we obtain
\begin{equation}\label{eq:8}
\displaystyle  \frac{\partial P}{\partial t} =\Gamma (1 - \wp) \left[ \frac{\partial}{\partial \nu} \nu  + \frac{\partial}{\partial \nu^*} \nu^*  \right] P + \frac{1}{T_1} \frac{\partial}{\partial \mu} \mu P + \frac{2\gamma_{\uparrow} (\Gamma + 2 \gamma_{\downarrow})}{(\gamma_{\uparrow}+\gamma_{\downarrow})} \frac{\partial^2}{\partial \nu \partial \nu^*}P.
\end{equation}
The equation can be solved by separation of variables, and we calculate the steady-state correlation functions as in statistical mechanics
\begin{eqnarray}\label{eq:9}
&\displaystyle \langle a^+ a \rangle_{ss,<} = \frac{g^2}{\kappa^2} \langle J_+ J_- \rangle_{ss,<} = \frac{Ng^2}{\kappa^2 (1 - \wp)} \frac{\gamma_{\uparrow} (\Gamma + 2 \gamma_{\downarrow})}{\Gamma (\gamma_{\uparrow}+\gamma_{\downarrow})},&\\
\label{eq:10}
&\displaystyle g^{(1)}_< (\tau) = \lim_{t \rightarrow \infty} \frac{\langle a^+(t) a(t+\tau) \rangle_<}{\langle a^+ a \rangle_{ss,<}} = e^{- \Gamma (1-\wp) \tau} e^{-i \omega \tau}, \quad \tau \gg 1/\kappa,&\\
\label{eq:11}
&\displaystyle g^{(2)}_<(\tau) = \lim_{t \rightarrow \infty} \frac{\langle a^+(t) a^+(t+\tau) a(t+\tau) a(t) \rangle_<}{\langle a^+ a \rangle_{ss,<}^2} = 1 + e^{-2 \Gamma (1-\wp) \tau}, \quad \tau \gg 1/\kappa.&
\end{eqnarray}

The result for two-time correlation functions is analogous to the case of
good-cavity laser, up to the replacement $\Gamma \rightarrow \kappa$, since we
adiabatically eliminate the cavity mode, rather than the polarization of
active atoms \cite[ch. 8.1.4]{Carmicael_1}. Note, that due to adiabatic mode
elimination we can only resolve times $\tau \gg 1/\kappa$. The smaller times
were resolved in paper \cite{Andrianov2013}, but only for the model with a single active atom.

In the case $\wp = 1$
the drift term in the Fokker-Planck equation (\ref{eq:8}) vanishes and there
is no restoring force to prevent the fluctuations from growing without bound.
Thus, the average number of plasmons in the cavity mode (\ref{eq:9}) diverges
at the point $\wp = 1$. Thereby, the equation (\ref{eq:8}) cannot correctly
describe the behaviour of system at the generation threshold.
Note, that the operation of a bad-cavity laser at the threshold was discussed in paper \cite{Gnutzmann1998}.

\section{Quantum Fluctuations Above Threshold}

Now, we turn out to the description of fluctuations above the generation threshold. As follows from the steady-state solution in the case $\wp > 1$, the phase of polarization is undetermined. Thus, in place of the first equation in (\ref{eq:4}), we write
\begin{equation}\label{eq:12}
\displaystyle v e^{i \omega t}/N =e^{i N^{-1/2} \psi} \left(|\sigma| + N^{-1/2} \nu \right),
\end{equation}
where now the variable $\nu$ represents real amplitude fluctuations, which can be both positive and negative, but must fall within the range $-N^{1/2} |\sigma| \leq \nu \leq \infty $, and the variable $\psi$ represents phase fluctuations. The distribution function in scaled variables, normalized with respect to the integration measure $d\nu d\psi d\mu$, is defined by $P(\nu,\psi,\mu,t) \equiv  N^{3/2} \left(|\sigma| + N^{-1/2} \nu \right) \tilde{P} \left(v(\nu,\psi,t),v^*(\nu,\psi,t),m(\mu,t),t \right)$.

In the case of small amplitude fluctuations above the generation threshold, $|\sigma| \gg N^{-1/2} \nu$, we can partially separate variables $P(\nu,\psi,\mu,t) = A(\nu,\mu,t) \Phi(\psi, t)$. Moreover, in accordance with experimental papers \cite{Trieschmann2011, Kim2012}, we believe $\Gamma T_1 \gg 1$. With these assumptions the evolution of the distribution functions $A$ and $\Phi$ are governed by equations:
\begin{eqnarray}
\label{eq:14}
&\displaystyle \frac{\partial A}{\partial t} = \sqrt{\frac{ \Gamma (\wp -1)}{4T_1} } \left[ 8 \frac{\partial}{\partial \mu} \nu - \frac{\partial}{\partial \nu} \mu \right]A+ \frac{1}{T_1} \frac{\partial}{\partial \mu} \mu A +  \frac{\gamma_p}{4} \left( 1+ \frac{1}{\wp_0}\right) \frac{\partial^2}{\partial \nu^2} A,&\\
\label{eq:15}
&\displaystyle \frac{\partial \Phi}{\partial t} = \gamma_p \frac{\Gamma T_1 (\wp_0+1) \wp_0}{(\wp-1)}  \frac{\partial^2}{\partial \psi^2} \Phi.&
\end{eqnarray}

Solving the equation (\ref{eq:14}), we can calculate the average number of plasmons in the cavity mode above the generation threshold
\begin{equation}\label{eq:16}
\displaystyle \langle a^+ a \rangle_{ss,>} - \frac{\Gamma (\wp-1)}{4 T_1 g^2}  =\frac{ \gamma_p (\wp_0+1)}{8 \kappa } \left[ 2 \Gamma T_1 + 1/(\wp-1) \right].
\end{equation}
Here the second term in the left part corresponds to the steady-state solution of macroscopic equations and the right part describes fluctuations. Our theory is correct if fluctuations are small. Far enough away from the threshold, when $\Gamma T_1 (\wp -1) \gg 1$, this leads to the restriction $\wp -1 \gg (\gamma_p / \kappa) (g T_1)^2 (\wp_0 +1)$. Thus, our theory is self-consistent if $(\gamma_p / \kappa) (g T_1)^2 \ll 1$.

\begin{figure}[tbp]
\centering\includegraphics[width=7cm]{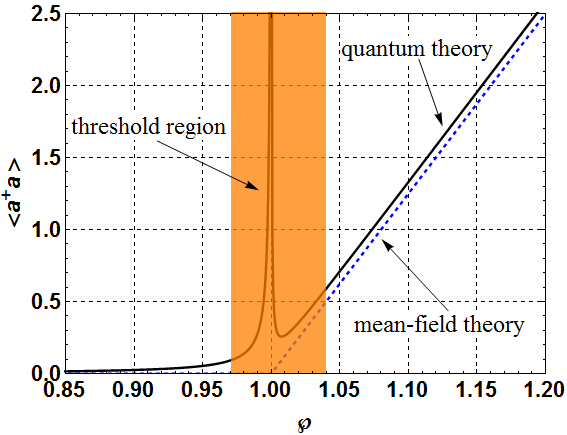}\\
\caption{The dependence of the average number of plasmons in the cavity on the pump-parameter $\wp$, for the following parameters: $\kappa = 2 \cdot 10^{15}\,s^{-1},\; \Gamma = 5 \cdot 10^{12}\,s^{-1},\; g = 10^{11}\,s^{-1},\; \gamma_{\uparrow} = 9 \cdot 10^{10}\,s^{-1},\; \gamma_{\downarrow} = 10^{10}\,s^{-1}$. The dashed line corresponds to the mean-field theory, the solid line takes into account quantum fluctuations.}\label{pic:1}
\end{figure}

Next, completely ignoring the small amplitude fluctuations in equation (\ref{eq:12}) and using equation (\ref{eq:15}), we can obtain fist-order correlation function ($\tau \gg 1/\kappa$)
\begin{equation}\label{eq:17}
g_>^{(1)} (\tau) = e^{-(i \omega+ D) \tau },\quad \displaystyle D = \gamma_p \frac{\Gamma T_1}{N} \frac{\wp_0 (\wp_0+1)}{(\wp-1)} =\frac{ \gamma_p \Gamma}{4} \hbar \omega \frac{(\wp_0 +1)}{P_{>}} \ll 1/T_1,
\end{equation}
where $D$ defines the width of spectral line above the generation threshold,
and we rewrite it in terms of the output power $P_{>} = \kappa
\hbar \omega \langle a^+ a \rangle_{ss,>}$. Qualitatively, the result is
similar to the case of good-cavity lasers, compare with
\cite[(8.138)]{Carmicael_1}. However, the mechanism which leads to the
narrowing of the spectral line is quite different. We will discuss it in
detail in the next section.

Finally, we obtain the second-order correlation function, taking into account the amplitude fluctuations in the main order. In the region of interest, $\Gamma T_1 (\wp -1) \gg 1$, one can find ($\tau \gg 1/\kappa$)
\begin{equation}\label{eq:18}
\displaystyle g_{>}^{(2)} (\tau) = 1 + 4 \gamma_p T_1 \frac{\Gamma T_1}{N} \frac{\wp_0 (\wp_0+1)}{(\wp-1)}    e^{- \tau /2 T_1} \cos \left( \sqrt{8 \Gamma T_1 (\wp - 1)} \tau/2 T_1 \right).
\end{equation}
The damped oscillations in $g_{>}^{(2)} (\tau)$  are usual for bad-cavity lasers \cite{Gnutzmann1998}.
The similar behaviour had been observed in numerical calculations in the paper \cite{Temnov2009a}, but there authors dealt with the regime of large fluctuations. As we discussed above, our consideration is reliable only in small-noise limit, i.e. $g_{>}^{(2)} (0) - 1 \ll 1$. Note, that in the case of good-cavity lasers the amplitude fluctuations does not lead to the oscillations in $g_{>}^{(2)} (\tau)$, see~\cite[(8.139)]{Carmicael_1}.
In the next section we will explain the origin of damped oscillations in detail and establish when the good-cavity behaviour is replaced by the bad-cavity damped oscillations.

\section{Numerical Parameters and Discussion}

\begin{figure}[tbp]
\centering\includegraphics[width=11cm]{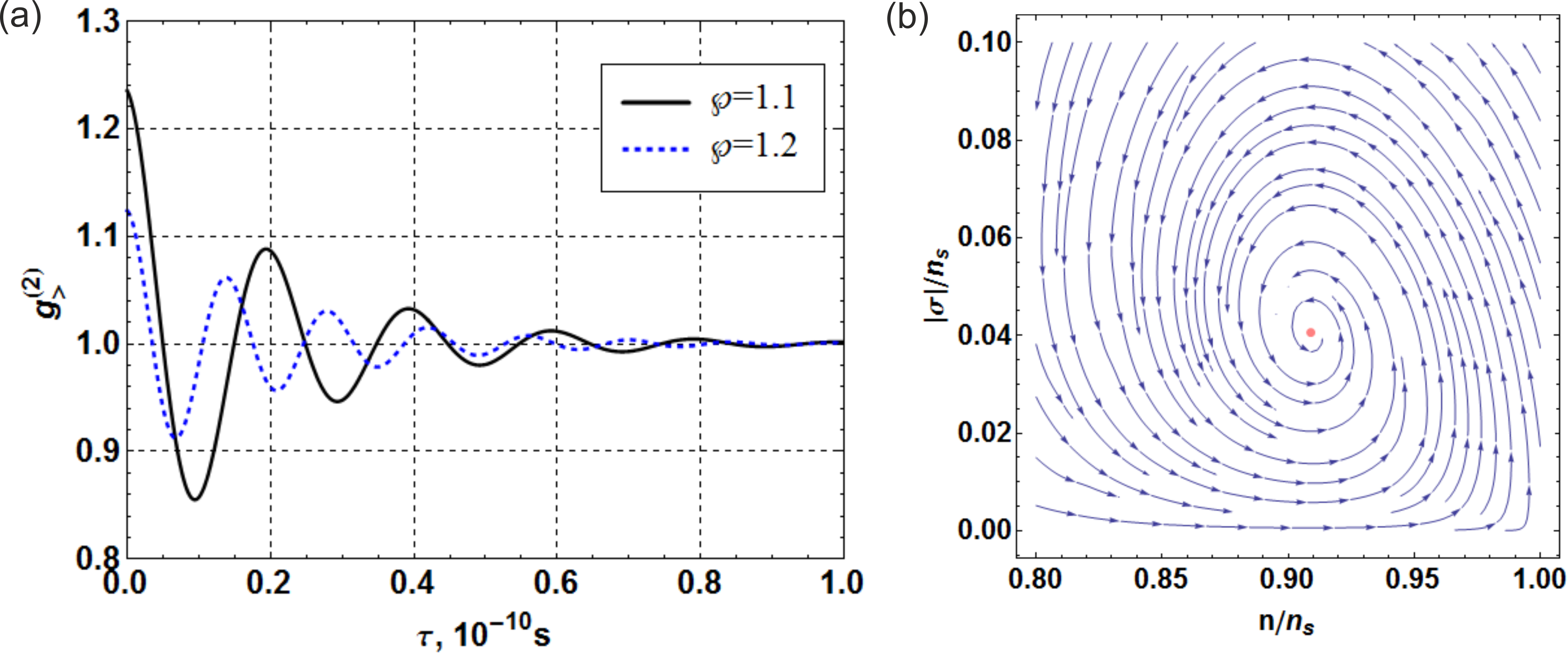}\\
\caption{(a) The second-order correlation function above the generation threshold. The parameters are as for the Fig.~\ref{pic:1}. (b) The vector field obtained from the right parts of the macroscopic equations with $\wp=1.1$, $n_s=0.8$ and $\Gamma T_1 = 50$. The spiral movement to the steady-state leads to the damped oscillations in $g^{(2)}_{>} (\tau)$.}\label{pic:2}
\end{figure}

In the Fig.~\ref{pic:1} we plot the dependence of the average number
of plasmons in the cavity on the pump-parameter $\wp$ below and above the
generation threshold, where we propose the following sets of parameters:
$\kappa = 2 \cdot 10^{15}\,s^{-1},\; \Gamma = 5 \cdot 10^{12}\,s^{-1},\; g =
10^{11}\,s^{-1},\; \gamma_{\uparrow} =9 \cdot 10^{10}\,s^{-1},\;
\gamma_{\downarrow} = 10^{10}\,s^{-1}$ and we variate the parameter $N$. The
dashed line corresponds to the semiclassical mean-field theory, see equations
(\ref{eq:5})--(\ref{eq:7}), and the solid line takes into account quantum
fluctuations, see equations (\ref{eq:9}), (\ref{eq:16}). Emphasize, that near
the threshold $\wp \sim 1$ the average number of plasmons in the cavity mode
$\langle a^+ a \rangle < 1$. Despite this fact, a linewidth of the order of
$\Gamma$ well below threshold is replaced by a much narrower line of the
order of $D$ above threshold. When $\wp-1 = 0.1$, we find $D/\Gamma \sim
1/850$.

In the good-cavity lasers the spectral line width becomes narrower above the
generation threshold because the stimulated emission starts to play more
important role than the spontaneous emission. In this case the average
number of photons in the cavity near the threshold is sufficiently greater
than unity \cite{Scully, Carmicael_1}. In our case the average number of
plasmons is less than unity. Thus, in order to spectral line width becomes
narrower, the born plasmon should be coherent with the already dead plasmon,
and this is possible since the active atoms preserve the coherence.
Originally arising plasmon interacts with active atoms and make them coherent
to each other. Then the plasmon dies after a short time $\sim 1/\kappa$, but
the coherence is still alive in active atoms, which relax slowly, $1/\Gamma
\gg 1/\kappa$. The next plasmon generated by such atoms can be coherent to
the previous one. This mechanism of the spectral line narrowing was
demonstrated in experiment with the laser, which deals with photons, in the
paper \cite{Bohnet2012a}. However, we cannot directly applied our theory to
this experiment, since the assumption $\Gamma T_1 \gg 1$ is not fulfilled.
Another experimental realization of the lasing regime is \cite{Noginov2009},
where mean number of plasmons is also less than unity, $\langle a^+ a \rangle \sim 0.2$. Indeed, the
total pumping energy absorbed per one nanolaser is $P_W \sim \langle a^+ a
\rangle \hbar \omega^2 \tau_p / Q$, where $\tau_p$ is the duration of the
pumping pulse. The measured value $P_W = 10^{-13} J$ near the generation
threshold, $Q=13.2$, $\tau_p = 5 ns$ and $\hbar \omega = 2.3 eV$.
Note, that numerical parameters in the Fig.~\ref{pic:1} are
slightly different from those from the paper~\cite{Noginov2009}. The reason
is that for the experimental parameters the amplitude fluctuations are large,
see (\ref{eq:16}), and our theory is not applicable well. However, we believe
that the picture does not change qualitatively.

Next, in the Fig.~\ref{pic:2}(a) we plot the second-order correlation function
above the generation threshold, according to the expression (\ref{eq:18}).
The dependence is valid if $\tau \gg 1/ \kappa$, because it is obtained under
the bad-cavity approximation.
In order to explain the nature of the damped oscillations, we plot the vector field on the $n$-$|\sigma|$-plane, see
Fig.~\ref{pic:2}(b), which corresponds to the right parts of the mean-field
equations (\ref{eq:5})-(\ref{eq:7}). The red point represents the
steady-state solution of these equations. Fluctuations move the system from
its equilibrium state and then it relaxes to the steady-state. The spiral
movement corresponds to the damped oscillations in polarization amplitude
$|\sigma|$ and inverse population $n$, and as a consequence in the second-order correlation function.

\begin{figure}[tbp]
\centering\includegraphics[width=7cm]{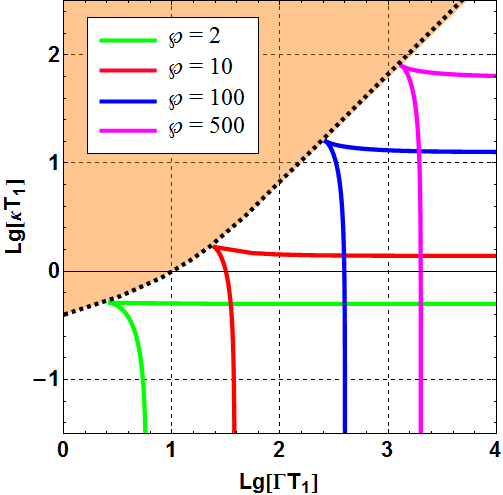}\\
\caption{The "phase diagram" contained information about oscillations in $g^{(2)}_{>} (\tau)$, for different pump-parameters $\wp$. The area below and to the right to the corresponding curves responds to the non-oscillating regime. The painted area above the dotted line corresponds to the bad-cavity lasers.}\label{pic:3}
\end{figure}

In the same way we can analyze a laser with a resonator of arbitrary $Q$-factor, but we should go beyond the bad-cavity approximation.
Instead of macroscopic equations (\ref{eq:5})-(\ref{eq:7}), we need to consider a full system of three Maxwell-Bloch equations \cite{Stockman2010}.
The evolution of amplitude fluctuations around the steady-state solution is defined by three eigenvalues. One eigenvalue is always real. Two others can be either real or complex conjugate, which corresponds to non-oscillating and oscillating character of the second-order correlation function respectively. The eigenvalues are fully defined by three parameters: $\kappa T_1,\; \Gamma T_1$ and $\wp$. In the Fig.~\ref{pic:3} we plot a "phase diagram" in logarithmic coordinates for different pump-parameters $\rho > 1$. The area to the right and below to the corresponding curves responds to the non-oscillating regime. The parameters from the painted area above the dotted line always correspond to the oscillations in $g^{(2)}_{>} (\tau)$. This is an area of bad-cavity lasers, where the mechanism of spectral line narrowing is based on the coherence conservation in the state of active atoms. The asymptotic behaviour ($\wp \gg 1$) of the dotted line was obtained numerically and it corresponds to the dependencies $\kappa T_1 = 0.16 \wp$ and $\Gamma T_1 = 2.5 \wp$. In the area below the dotted line both the oscillating and non-oscillating behaviour of $g^{(2)}_{>} (\tau)$ is possible, depending on pump-parameter $\wp$. Thus, the shape of the second-order correlation function provides insufficient information to obtain a mechanism of spectral line narrowing. The answer on this question is contained in the "phase diagram" in the Fig.~\ref{pic:3}.

\section{Conclusion}

To summarize, the quantum theory of a spaser-based nanolaser was presented. We found that the average number of plasmons in the cavity mode near the generation threshold can be less than unity both in our theory and experiments~\cite{Noginov2009}. Despite this fact, the spectral line width narrows sufficiently, when passing through the threshold. We proposed that it is possible behaviour when the coherence preserved by the active atoms, which relax slowly than the damping of cavity mode occurs. We also studied the amplitude fluctuations of polarization and concluded that they lead to the damped oscillations in the second-order correlation function $g^{(2)} (\tau)$ above the generation threshold. It is unusual behaviour for the good-cavity lasers, and we investigated in detail at what relationship between cavity decay rate $\kappa$ and homogeneous broadening of active atoms $\Gamma$ the bad-cavity damped oscillations are replaced by non-oscillating regime. In the future, we plan to extend our consideration beyond to the small-noise limit.

\section{Acknowledgments}

We thank V.V. Lebedev and V.P. Drachev for fruitful discussions. The work was supported by GRANT XXX.

\end{document}